\documentstyle[letter,mbf,epsfig,wrapft]{ptptex}

\def\bge{\begin{equation}}
\def\ene{\end{equation}}
\def\bg{\begin{eqnarray}}
\def\en{\end{eqnarray}}

\title{Relationship between Quark-Meson Coupling Model and
Quantum Hadrodynamics}

\author{Koichi Saito}

\inst{Tohoku College of Pharmacy, Sendai 981-8558, Japan}

\recdate{
}

\abst{
Using the quark-meson coupling (QMC) model, we study nuclear matter from
the point of view of quark degrees of freedom.  Performing a
re-definition of the scalar field in matter, we transform QMC to a
QHD-type model with a non-linear scalar potential.  The potentials
obtained from QMC are compared with those of the relativistic
mean-field models.  
}

\begin{document}

\maketitle

Since at present rigorous studies of quantum chromodynamics (QCD) are
limited to matter system with high temperature and zero baryon
density, it is important to build models which help to bridge the
discrepancy between nuclear phenomenology and QCD.  We have proposed a
relativistic quark model for nuclear matter and finite nuclei, that 
consists of non-overlapping nucleon bags bound by the self-consistent
exchange of isoscalar, scalar ($\sigma$) and vector ($\omega$) mesons
in mean-field approximation (MFA) -- this model is called the
quark-meson coupling (QMC) model.\cite{qmc}  
On the other hand, recent theoretical studies show that various properties
of finite nuclei can be very well described by the relativistic
mean-field (RMF) models, i.e., Quantum hadrodynamics (QHD).\cite{qhd}  
In this letter, we consider relationship between QMC and QHD and study
how the internal structure of the nucleon sheds its effect on
effective nuclear models. 

In our previous works,\cite{qmc} the MIT bag model has been used to
describe the quark structure of the nucleon. 
Since the confined quarks interact with the scalar field, $\sigma$, in 
matter, the effective nucleon mass, $M_N^\star$, in QMC is given by a
function of $\sigma$ through the quark model of the nucleon. (Although
the quarks also interact with the $\omega$ meson, it has no effect on
the nucleon structure except for a shift in the nucleon
energy.\cite{qmc})  
The (relativistic) constituent quark model (CQM) is an alternative
model for the nucleon.  Recently, Shen and Toki\cite{toki} have
proposed a new version of QMC -- the quark mean-field (QMF) model,
where CQM is used to describe the nucleon.  

In the present study, as well as the bag model (BM), we want to use
the relativistic CQM with confining potentials,
$V(r)$, of a square well (SW) and a harmonic oscillator (HO) to see the
dependence of the matter properties on the quark model.  It is 
assumed that the light ($u$ or $d$) quark mass, $m_q$, is 300 MeV in
CQM, while $m_q=0$ MeV in
BM. Furthermore, we introduce a Lorentz-vector type confining
potential, which is proportional to $\gamma_0$, as well as the scalar
one:  
\begin{equation}
V(r) = (1 + \beta \gamma_0) U(r),
\label{potential}
\end{equation}
where the potential, $U(r)$, is given by SW or HO and $\beta (0 \leq
\beta < 1)$ is a parameter to control the strength of the
Lorentz-vector type potential.  We assume that the shape of the
Lorentz-vector type confining potential is the same as that of the
scalar type one. 

In SW, the solution for a quark field, $\psi_q$, can be calculated
{\it {\`a} la Bogolioubov}.\cite{bogo}  The potential is given by
$U(r) = 0$ for $r \leq R$ and $M$ for $\ r > R$, where $R$ is the
radius of the spherical well and $M$ is the height of the potential
outside the well.  After finishing all calculations, the limit $M
\to \infty$ is taken.\cite{bag}  This system may be described by 
Lagrangian density
\begin{equation}
{\cal L}_{SW} = {\bar \psi}_q (i\gamma\cdot\partial - m_q) \psi_q
\theta(R-r) - \frac{1}{2}{\bar \psi}_q (1+\beta \gamma\cdot a)
\psi_q \delta(r-R),
\label{swlag}
\end{equation}
where $a^\mu$ is the unit vector in time direction: $a^\mu =
(1, {\vec 0})$.  This Lagrangian provides a boundary condition: $i
\gamma\cdot n \psi_q = (1+\beta \gamma\cdot a) \psi_q$ at
$r=R$, where $n^\mu$ is the unit normal outward from the potential
surface. This condition gives 
\begin{equation}
j_0(x) = \sqrt{\frac{(1-\beta)(E-m_q)}{(1+\beta)(E+m_q)}} j_1(x) 
\ \ \mbox{at} \ r=R, 
\label{bound}
\end{equation}
with $j_n$ the spherical Bessel function, $x$ the eigenvalue 
of the confined quark and $E$ the quark
energy.  (Note that in fact the quark current flows in only the
azimuthal direction.\cite{saito})
To take into account the corrections of the spurious center of mass
(c.m.) motion and gluon fluctuations, we add the familiar form,
$-z/R$, with a parameter $z$ for those corrections to the total 
energy.\cite{bogo,bag}  In SW, the nucleon mass, $M_N$, (at rest) is
then given by $M_N = (3 \alpha - z)/R$, where $\alpha^2 =
x^2+\xi^2$ and $\xi = Rm_q$.  

In case of HO, we set $\beta = 1$ in Eq.(\ref{potential}) because the
quark wave function can be obtained analytically.\cite{ho}  When $U(r) =
\frac{1}{2} c r^2$ ($c$ the oscillator strength), a condition to
determine the quark energy is given by
\begin{equation}
\sqrt{E+m_q} (E-m_q) = 3 \sqrt{c}.  
\label{hocond}
\end{equation}
The c.m. energy can be evaluated exactly, as in the non-relativistic
harmonic oscillator, and it is just one third of the total
energy.\cite{ho}  Thus, the nucleon mass is given by $M_N = 2E - E_g$,
where $E_g$ describes gluon fluctuation corrections.\cite{ho} 

{}For the MIT bag model there are many good reviews.\cite{bag}  In BM,
we take $m_q=0$ MeV and $\beta=0$. (Even in BM it is possible to
include the Lorentz-vector type potential using Eqs.(\ref{swlag}) and
(\ref{bound}).  However, if we use a large $\beta$ in BM, it is hard to get
good values of the nuclear matter properties.)  

Now we consider an iso-symmetric nuclear matter with Fermi momentum
$k_F$, which is given by $\rho_B = 2 k_F^3 / 3\pi^2$ ($\rho_B$ the
nuclear matter density).  Then, the total energy per nucleon,
$E_{tot}$, can be written as\cite{qmc}
\begin{equation}
E_{tot} = \frac{4}{\rho_B (2\pi)^3} \int^{k_F}
d\vec{k} \sqrt{M_N^{\star 2}(\sigma) + \vec{k}^2} + \frac{m_{\sigma}^{2}}
{2\rho_B}{\sigma}^2 + \frac{g_{\omega}^2}
{2m_{\omega}^{2}}\rho_B ,
\label{tote}
\end{equation}
where $M_N^\star$ is calculated
by the quark model.  The $\sigma$ and $\omega$ meson masses, $m_\sigma$ and
$m_\omega$, are taken to be 550 MeV and 783 MeV, respectively.  
The $\omega$ field is determined by baryon number conservation:
$\omega =g_{\omega}\rho_B / m_{\omega}^{2}$ ($g_\omega$ is the
$\omega$-nucleon coupling constant), while the scalar mean-field is
given by a self-consistency condition: $(\partial E_{tot} /
\partial \sigma) = 0$.\cite{qmc} 

In SW, we set the radius of the potential to be $R = 0.8$ fm and
detrmine $z$ so as to fit the free nucleon mass, $M_N$ (= 939 MeV).
The parameter $\beta$ is chosen to be 0 and 0.5 to examine the effect
of the Lorentz-vector type confining potential.  We find that
$z=4.396$ and $5.164$ for $\beta = 0$ and $0.5$, respectively. In 
HO, there are two adjustable parameters, $c$ and $E_g$.  We
determine those parameters so as to fit the free nucleon mass and the
root-mean-square (charge) radius of the free proton: $r_N^2 =
0.6$ fm$^2$.\cite{rms} ($r_N$ is calculated by the quark wave
function.)  We find that $c = 1.591$ fm$^{-3}$ and $E_g = 344.7$ MeV
for the free nucleon.  In nuclear matter, we keep $c$ and $E_g$
constant and the quark energy $E$ varies, depending on the scalar
field.  In BM, the bag constant, $B$, and the parameter, $z$, are
fixed to reproduce the free nucleon mass.  As in SW, we choose the bag
radius of the free nucleon to be 0.8 fm.  We find $B^{1/4}$ = 170.3
MeV and $z = 3.273$.\cite{qmc}  

\begin{wraptable}{l}{\halftext}
\caption{Coupling constants, $M_N^\star$ and $K$.  The effective
nucleon mass, $M_N^\star$, is calculated at $\rho_0$.  The nuclear
incompressibility, $K$, is quoted in MeV. The SW model with $\beta =
0(0.5)$ is denoted by SW0(5). 
}
\label{tab:cc}
\begin{center}
\begin{tabular}{ccccc} \hline \hline
 & $g_\sigma^2$  & $g_\omega^2$ & $M_N^\star/M_N$ & $K$ \\ \hline
SW0 & 84.4 & 104 & 0.725 & 329 \\
SW5 & 66.6 & 65.2 & 0.807 & 287 \\
HO & 147 & 64.5 & 0.805 & 309 \\
BM & 67.6 & 66.1 & 0.805 & 278 \\ \hline
\end{tabular}
\end{center}
\end{wraptable}

Now we are in a position to determine the coupling constants: 
the $\sigma$-nucleon coupling constant, $g_{\sigma}^2$, and
$g_{\omega}^2$ are fixed to fit the nuclear binding energy ($-15.7$
MeV) at the saturation density ($\rho_0 = 0.15$ fm$^{-3}$) for
nuclear matter.  The coupling constants and some calculated properties
for matter are listed in Table~\ref{tab:cc}.  The present
quark models can provide good values of the nuclear incompressibility,
$K$.  

In SW and BM with {\em massless} quarks, the quark scalar density in
the nucleon\cite{qmc} vanishes in the limit $\beta \to 1$, which means
that the $\sigma$ meson does not couple to the nucleon.\cite{ks}  
This fact implies that as $\beta$ is larger the $\sigma$-nucleon
coupling is weaker in matter. Thus, we can conclude that qualitatively
a large mixture of the Lorentz-vector type confining potential leads
to a weak scalar mean-field and hence a large effective nucleon mass
in nuclear matter.  Since in MFA a small effective nucleon mass (and
hence a strong scalar field) is favorable to fit various properties
of finite nuclei,\cite{qhd} the confining potential including a strong
Lorentz-vector type one may not be suitable for describing a nuclear
system.   

The main difference between QMC and QHD at {\em hadronic}
level\cite{qmc} lies in the dependence of the nucleon mass on the
scalar field in matter.  By performing a re-definition of the scalar
field, the QMC Lagrangian density\cite{qmc} can be cast into a form
similar to a QHD-type mean-field model, in which the nucleon mass
depends on the scalar field linearly, with self-interactions of the
scalar field.\cite{muller} 
In QMC, the nucleon mass in matter is given by a function
of $\sigma$, $M_{N,QMC}^{\star}(\sigma)$, through the quark model of
the nucleon, while in QHD the mass depends on a scalar
field linearly, $M_{N,QHD}^{\star} = M_{N} - g_0 \phi$ ($\phi$ is the
scalar field in a QHD-type model).  Hence, to transform QMC into a
QHD-type model, we can apply a re-definition of the scalar field,  
\begin{equation}
g_0 \phi(\sigma) = M_N - M_{N,QMC}^{\star}(\sigma) ,
\label{redef}
\end{equation}
to QMC, where $g_0$ is a constant chosen so as to normalize the scalar
field $\phi$ in the limit $\sigma \to 0$: $\phi(\sigma) = \sigma +
{\cal O}(\sigma^2)$. Thus, $g_0$ is given by 
$g_0 = - ( \partial M_{N,QMC}^{\star}/\partial \sigma )_{\sigma = 0}$. 
In QMC, we find $g_0 = g_\sigma$ for SW and BM, while $g_0 =
\frac{2}{3}g_\sigma$ for HO. 

The contribution of the scalar field to the total energy, $E_{scl}$, is now 
rewritten in terms of the new field $\phi$ 
\begin{equation}
E_{scl} = \frac{1}{2} \int d\vec{r} \ [(\nabla \sigma)^2 + m_{\sigma}^2
\sigma^2 ] 
= \int d\vec{r} \ \left[ \frac{1}{2}h(\phi)^2(\nabla \phi)^2 +
U_s(\phi) 
\right] ,
\label{es}
\end{equation}
where $U_s$ describes the self-interactions of the scalar field
\begin{equation}
U_s(\phi) = \frac{1}{2} m_{\sigma}^2 \sigma(\phi)^2 \ \ \mbox{and} \ \
h(\phi) = \left( \frac{\partial \sigma}{\partial \phi} \right)
= \frac{1}{m_\sigma \sqrt{2U_s(\phi)}}
\left( \frac{\partial U_s(\phi)}{\partial \phi} \right) .
\label{Us}
\end{equation}
Note that in uniformly distributed nuclear matter the derivative
term in $E_{scl}$ does not contribute.  
(The effect of this term on the properties of finite nuclei has been 
studied in Ref.~\cite{muller}.) 
Now QMC can be re-formulated in terms of the new scalar
field, $\phi$, and it is of the same form as QHD with the
non-linear scalar potential, $U_s(\phi)$, and the coupling, $h(\phi)$,
to the gradient of the scalar field.  (Note that since this re-definition of
the scalar field does not concern the vector interaction, the energy of
the $\omega$ field (see Eq.(\ref{tote})) is not modified.) 

The Zimanyi-Moszkowski (ZM) model\cite{zm} is a good example.  By
re-definig the scalar field, ZM can be {\em exactly} transformed to
a QHD-type model with a non-linear potential. 
Since the effective nucleon mass in ZM is given by\cite{zm}
\begin{equation}
M_{N,ZM}^\star = \frac{M_N}{1+(g_\sigma \sigma /M_N)},
\label{zmmass}
\end{equation}
the model involves higher order couplings between the $\sigma$ and the
nucleon.  Introducing a new scalar field $\phi$ by $g_0 \phi(\sigma) =
M_N - M_{N,ZM}^{\star}(\sigma)$, we easily find
\begin{equation}
g_0 = g_\sigma, \ \ 
\phi(\sigma) = \frac{\sigma}{1 + g^\prime \sigma} \ \ \mbox{and} \ \ 
\sigma(\phi) = \frac{\phi}{1 - g^\prime \phi} ,
\label{zmphi}
\end{equation}
with $g^\prime = g_\sigma /M_N$.  Thus, the non-linear potential is
given by Eq.(\ref{Us}) 
\begin{equation}
U_s(\phi) = \frac{1}{2} m_{\sigma}^2 \left( \frac{\phi}{1 - g^\prime
\phi} \right)^2 . 
\label{zmUs}
\end{equation}

In general, the in-medium nucleon mass may be given by a complicated
function of the scalar field.  However, in QMC the mass can be
parametrized by a simple expression up to
${\cal O}(g_\sigma^2)$:\cite{qmc} 
\begin{equation}
M_N^\star / M_N \simeq 1 - a y + b y^2 ,
\label{mparam}
\end{equation}
with $y (= g_\sigma \sigma / M_N)$ a dimensionless scale and 
two (dimensionless) parameters, $a$ and $b$. This parametrization is 
accurate up to $\sim 4\rho_0$.\cite{qmc}  

Once the parameters, $a$ and $b$, are fixed, we can easily re-define the
scalar field using Eq.(\ref{redef}).   We find 
\begin{equation}
g_0 = a g_\sigma, \ \ 
\phi(\sigma) = \sigma - d \sigma^2 \ \ \mbox{and} \ \ 
\sigma(\phi) = \frac{1-\sqrt{1-4d\phi}}{2d} ,
\label{qmcphi}
\end{equation}
with $d = bg_\sigma /aM_N$.  This satisfies the condition: $\sigma \to
0$ in the limit $\phi \to 0$.  The non-linear potential is thus calculated
\begin{eqnarray}
U_s(\phi) &=& \frac{m_\sigma^2}{2} \left( \frac{\sigma(\phi) - \phi}{d}
\right) , \nonumber \\
&=& \frac{m_\sigma^2}{2} \phi^2
+ g_\sigma r \left( \frac{m_\sigma^2}{M_N} \right) \phi^3
+ \frac{5}{2} g_\sigma^2 r^2 \left( \frac{m_\sigma}{M_N} \right)^2
\phi^4 + {\cal O}(g_\sigma^3) ,
\label{qmcnonlex}
\end{eqnarray}
where $r = b/a$.

\begin{wraptable}{l}{\halftext}
\caption{Parameters $a$, $b$, $\kappa$ and $\lambda$.
}
\label{tab:ab}
\begin{center}
\begin{tabular}{ccccc} \hline \hline
 & $a$  & $b$ & $\kappa$(fm$^{-1}$) & $\lambda$ \\
\hline
SW0 & 1.01 & 0.215 & 19.2 & 78.8 \\
SW5 & 1.02 & 0.497 & 39.0 & 327 \\
HO & 0.687 & 0.245 & 42.3 & 384 \\
BM & 0.998 & 0.435 & 35.1 & 264 \\ \hline
\end{tabular}
\end{center}
\end{wraptable}

The standard form of the non-linear scalar potential is usually given
by
\begin{equation}
U_s(\phi) = \frac{1}{2} m_{\sigma}^2 \phi^2 + \frac{\kappa}{6} \phi^3
+ \frac{\lambda}{24} \phi^4 .
\label{stUs}
\end{equation}
In more sophisticated version of QHD,\cite{newqhd} inspired by modern
methods of effective field theory, many other terms of meson-meson and
meson-nucleon couplings are considered.  However, we here focus on
only the parameters $\kappa$ and $\lambda$ to make our discussion simple.
It is well known that the non-linear scalar potential, Eq.(\ref{stUs}),
is practically indispensable to reproduce the bulk properties of finite
nuclei and nuclear matter in RMF. 

We can estimate the parameters, $\kappa$ and $\lambda$, in QMC by
comparing Eqs.(\ref{qmcnonlex}) and (\ref{stUs}). In
Table~\ref{tab:ab} the parameters, $a$, $b$, $\kappa$ and $\lambda$,
are presented.  The QMC model leads to the
non-linear potential with $\kappa \sim 20 - 40$ (fm$^{-1}$) and
$\lambda \sim 80 - 400$.  Since QMC predicts that both $a$ and $b$ are
{\em always positive},\cite{qmc,ks} we can expect that the quark
substructure of the in-medium nucleon provides a non-linear potential
with {\em positive} $\kappa$ and {\em positive} $\lambda$ in the
QHD-type mean-field model.  

\begin{figure}[t]
\parbox{\halftext}{
\epsfig{file=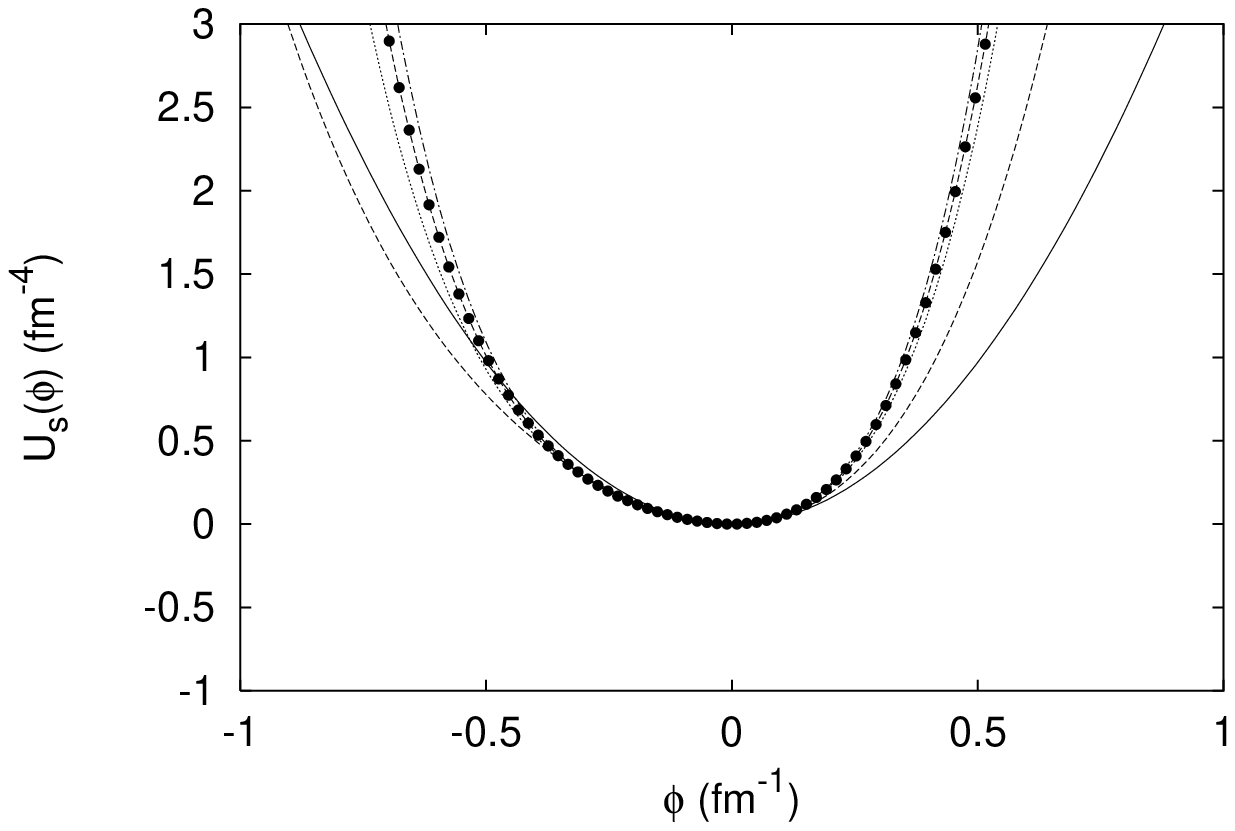,width=6.6cm}
\caption{Non-linear scalar potentials generated by QMC.
The solid curve shows $U_s = \frac{m_\sigma^2}{2} \phi^2$.
The dashed curve (with solid circles) is for SW
with $\beta = 0 (0.5)$, while the result of BM is
shown by the dotted curve.  The dot-dashed curve is for HO.
}
\label{fig:potqmc}}
\hspace{5mm}
\parbox{\halftext}{
\epsfig{file=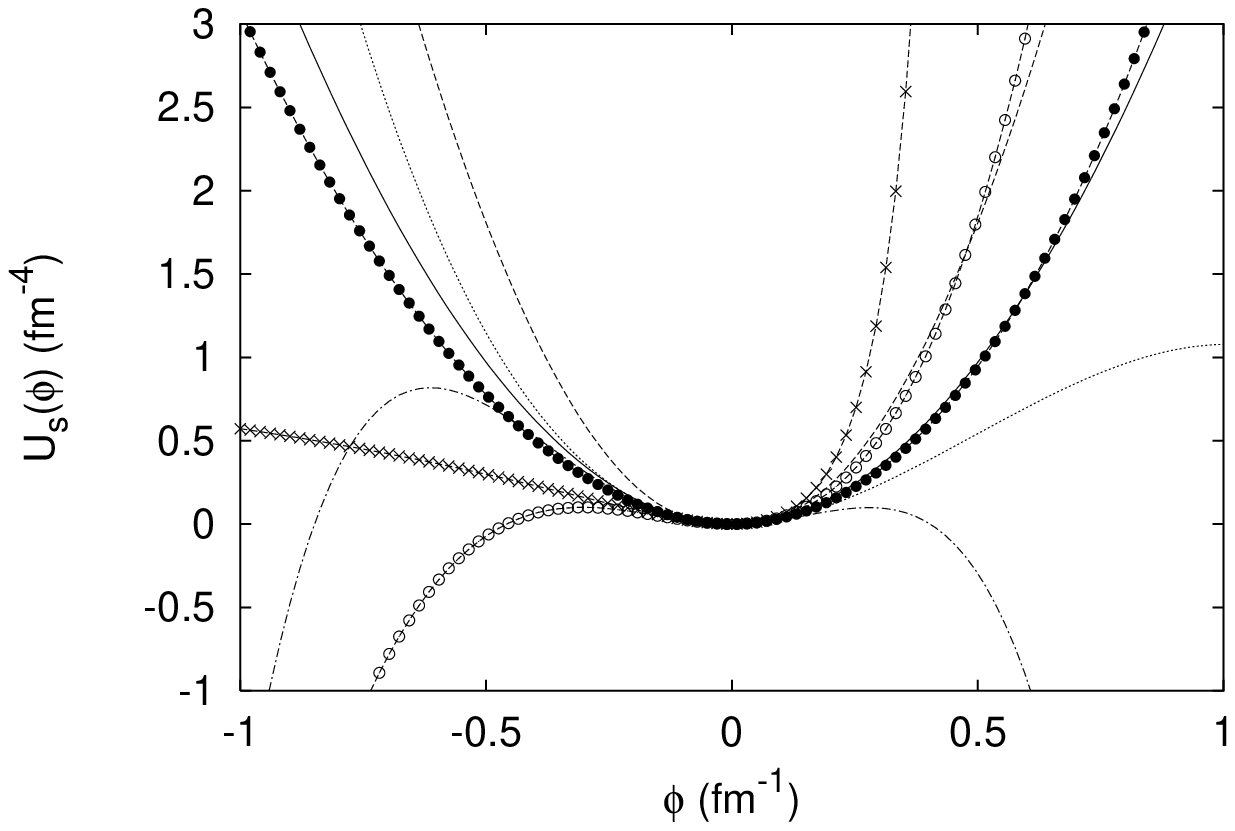,width=6.6cm}
\caption{Non-linear scalar potentials in RMF.
The solid curve shows $U_s = \frac{m_\sigma^2}{2} \phi^2$.
The dashed curve with open (solid) circles is for G2
(NLB),\protect\cite{newqhd} while the dashed one without any
marks is for PM3.\protect\cite{tm}  The dashed curve with
crosses is for ZM.\protect\cite{zm}  The potentials in
TM1\protect\cite{tm1} and NL1\protect\cite{nl1} are
respectively shown by the dotted and dot-dashed curves.
}
\label{fig:potrmf}}
\end{figure}

Using those parameters, we can draw the non-linear scalar potential 
generated by QMC, which is illustrated in Fig.~\ref{fig:potqmc}.  The
non-linear potentials, which have been used in various relativistic
mean-field models,\cite{zm,newqhd,tm,tm1,nl1} are also presented in
Fig.~\ref{fig:potrmf}.  From Fig.~\ref{fig:potqmc} we can see that the
quark models lead to the similar non-linear potentials, in spite of 
the big difference in the confinement mechanism.  On the other hand, 
the non-linear potentials in RMF show quite different
behaviors.  The potentials in NLB\cite{newqhd} and PM3\cite{tm} are 
relatively close to those produced by QMC. 

In summary, we have calculated the properties of nuclear matter using
QMC with various quark models for the nucleon.  Then, we have
performed a re-definition of the scalar field in matter and
transformed QMC to a QHD-type model with a non-linear scalar
potential.  The QMC model gives $\kappa \sim 20 - 40$ (fm$^{-1}$) and
$\lambda \sim 80 - 400$ for the non-linear scalar potential. The
shapes of the potentials generated by the quark models are very
similar to one another, although the confinement mechanism is quite
different in each model.  On the contrary, the parameters, $\kappa$
and $\lambda$, phenomenologically determined in RMF take various
values and the potentials in RMF are quite different from one another
for large $|\phi|$. In general, the phenomenological potential may
consist of the part, which is caused by the quark substructure of the
nucleon, and inherent self-couplings of the scalar field in matter.
It is very intriguing if the potential due to the internal structure
of the nucleon could be inferred by analyzing experimental data in the
future.

\end{document}